# Frameworks for Querying Databases Using Natural Language: A Literature Review

Hafsa Shareef Dar
Dept. of Software Engineering, University of Gujrat, 50700 Punjab, Pakistan
Email: hafsa.dar@uog.edu.pk

M. Ikramullah Lali, Moin Ul Din
Dept. of Computer Science, University of Gujrat, 50700 Punjab, Pakistan
Email: ikramlali@gmail.com; moinmalik007@yahoo.com

Khalid Mahmood Malik
Computer Science and Engineering Department, Oakland University, 2200 N. Squirrel Rd,
Rochester, MI 48309, USA
Email: mahmood@oakland.edu

Syed Ahmad Chan Bukhari*
Division of Computer Science, Mathematics and Science, College of Professional Studies, St. John's University,
New York, USA
Corresponding should be addressed to Syed Ahmed Chan Bukhari bukharis@stjohns.edu*

**Abstract:** A Natural Language Interface (NLI) facilitates users to pose queries to retrieve information from a database without using any artificial language such as the Structured Query Language (SQL). Several applications in various domains including healthcare, customer support and search engines, require elaborating structured data having information on text. Moreover, many issues have been explored including configuration complexity, processing of intensive algorithms, and popularity of relational databases, due to which translating natural language to database query has become a secondary area of investigation. The emerging trend of querying systems and speech-enabled interfaces revived natural language to database queries research area., The last survey published on this topic was six years ago in 2013. To best of our knowledge, there is no recent study found which discusses the current state of the art translations frameworks for natural language for structured and non-structured query languages. In this paper, we have reviewed 47 frameworks from 2008 to 2018. Out of 47, 35 were closely relevant to our work. SQL based frameworks have been categorized as statistical, symbolic and connectionist approaches. Whereas, NoSQL based frameworks have been categorized as semantic matching and pattern matching. These frameworks are then reviewed based on their supporting language, scheme of their heuristic rule, interoperability support, dataset scope and their overall performance score. The findings stated that 70% of the work in natural language to database querying has been carried out for SQL, and NoSQL share 15%, 10% and 5% of languages like SPAROL, CYPHER and GREMLIN respectively.  It has also been observed that most of the frameworks support English language only.

1. Introduction

Natural language to database querying frameworks translate natural language questions to valid database query languages. This translation helps to bridge the communication gap between non-technical users and database systems, as users do not require to understand the database schemas and query language syntax (Reis, 1997; Christian, 2010). Therefore, it is always desirable for the non-technical users to have a natural language interface for database querying. The history of natural language interface to database querying dates back to 1970s  when the LUNAR and LADDER systems were developed for non-technical users to pose natural language questions about the moon rock samples and US naval ships respectively (Woods, 1972). The rapid evolution of computer hardware and software in the last five decades have influenced databases in such a way that the database systems which were developed in 1970s are not even compatible with the current definition of a database (Bercich 2003; Frank 2018). Since then, several natural languages to database querying frameworks have been developed to fulfill the industry needs. By studying the development timeline of such systems, we have identified interesting research trends in translating natural language to database queries domain. The CHAT-80 was the leading natural language to database query system which was developed in 1980 (Warren, and Pereira, 1982). Early developed system had poor retrieval time, less support for the language portability, and had complex configuration processes. These factors contributed towards less adaptation of such systems for the commercial purposes.

Translating a natural language question into various database query languages such as SQL, Simple Protocol and RDF Query Language (SPARQL) is not a trivial task, as the current databases are diverse, gigantic in size and follow sophisticated data storage mechanisms (Nadkarni, 2011). Storage engines often store data in a variety of ways such as in structured format (tabular), No SQL or graph (text) or in hybrid format. Therefore, underlying storage engines require different query languages to retrieve the stored data. This heterogeneity of data storage mechanisms increases the complexity of natural language to database query translation. With the advancement of machine learning techniques, various frameworks have been developed and are able to efficiently translate natural language questions (from simple to complex questions) into database specific queries (SQL, NoSQL) (Yossi Shani, 2016) (Elías Andrawos, 2013).

The last review paper about natural language to database framework was published in 2013 (Sripad and n.d. 2013) which has classified the natural language querying framework for SQL only. Available review paper on this topic (Androutsopoulos, Ritchie and Thanisch, 1995) have mainly covered natural language to SQL database and highlighted the usage of developed systems so far. In this survey paper, we have reviewed Natural language to database querying frameworks developed for both the structured (SQL) and non-structured database query languages (NoSQL, GraphDB). Using Google Scholar, we have found thirty-five relevant frameworks published from 2008 to 2018. This review excludes papers which describe proposed approaches without corresponding evaluation i.e. precision and accuracy, on any benchmark. We have sub-divided the developed frameworks into two main categories (SQL and NoSQL) and provided a comprehensive review of each section (Figure 1). Moreover, for each category, a feature comparison among the developed frameworks documenting their salient features and highlighting their shortcomings has also been provided. The comparison has been conducted on different factors including language and approach supported, performance evaluation and others.

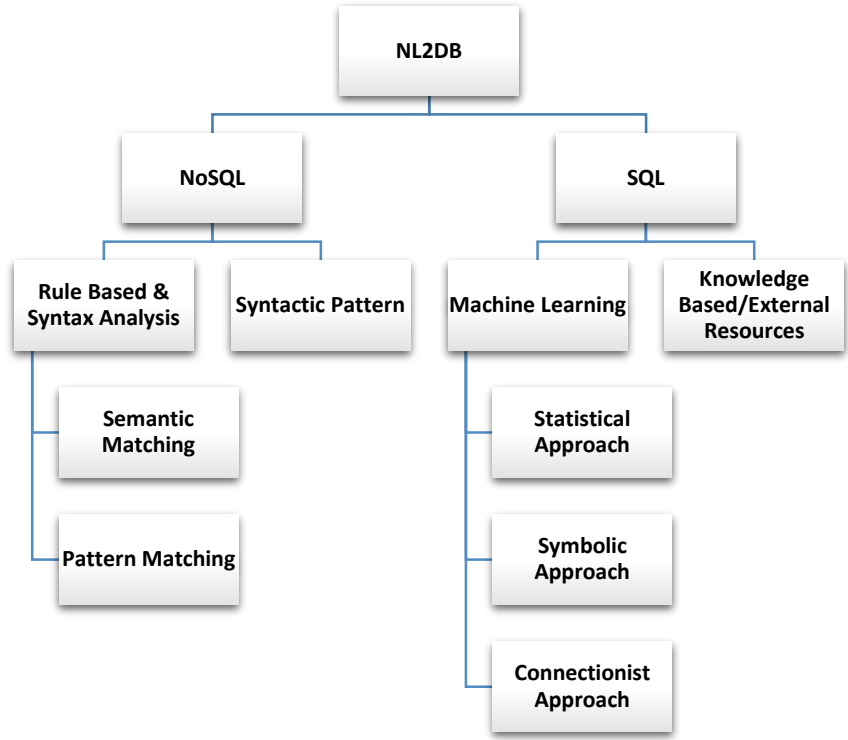

**Figure. 1 Classification of natural language to database querying frameworks.**

SQL and non-SQL categories can be further divided into rule based and syntax analysis, syntactic pattern, machine learning and knowledge based/external resources. Furthermore, these sub-categories have been reviewed for different approaches including semantic matching, pattern matching, supervised and unsupervised learning and statistical approach. Statistical approaches use large text corpora and perform analysis based on text characteristics without considering significant linguistic knowledge. Similarly, symbolic approach is widely used as a learning measures to different machine learning techniques. Connectionist approach proves to be an efficient model of learning tasks, therefore, the combination of connectionist with statistical or symbolic approach is an important area in natural language processing (Stefan, Ellen and Gabriele, 1996). Next section covers materials and methods used in conduction of this study.

## 1.1 Material and Methods

The most crucial part of this study was availability of relevant material. The articles were searched using authentic scientific databases including SPRINGER Link, IEEE, ACM Digital Library, Google Scholar, Emerald, Science Direct and Elsevier. Furthermore, some other databases were also explored but due to accessibility restrictions, they were not included. Search strategy was also designed based on different keywords like 'querying databases', 'natural language databases', 'frameworks for NLDB', 'natural language interfaces', 'SQL-based frameworks', and 'NoSQL frameworks'. Figure 2 further explains the selection procedure and keywords searching designed for this study.

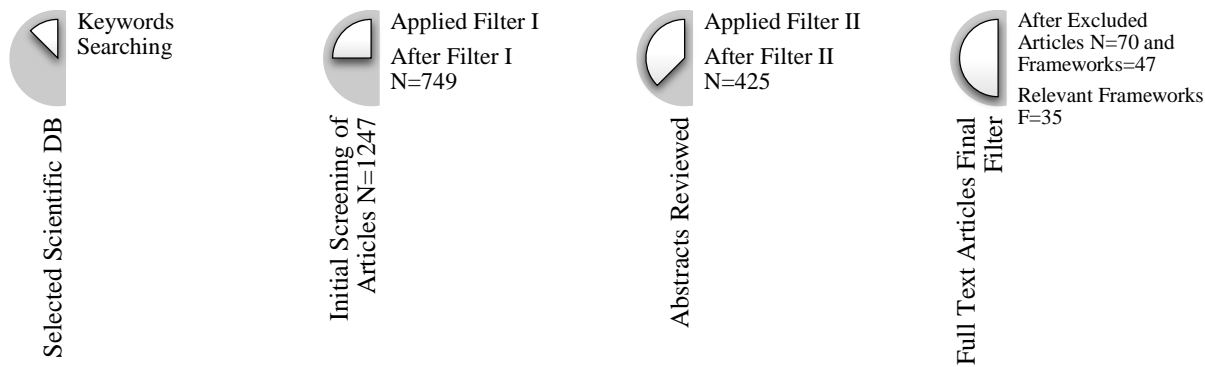

**Figure 2 Selection procedure and keyword search**

Figure 2 shows the selection procedure of articles and keywords searching applied in this study. For selection of articles, scientific databases were selected at first. This step helped to design search strategy for extracting more relevant material. In step two, initial screening was performed based on step one and 1247 articles were gathered. The articles were selected based on their titles and filtering was performed. After first filter, 749 articles seemed to be relevant. In step three, abstracts of the selected articles were studied, filtered and 425 articles were selected. After reviewing full text articles, 70 were selected because these articles have discussed 35 frameworks that are relevant to our work.

## 2. Background

This section presents a comprehensive review of the frameworks, shown in figure 1, that are developed for the natural language querying of structured (SQL) and unstructured (NoSQL) databases. A brief overview of these frameworks along with comparison of their features have been presented in this section.

### 2.1 Translation of the Natural Language to the Structured Query Language (SQL) Frameworks

These frameworks have been categorized into three different approaches namely statistical, symbolic and connectionist in machine learning and knowledge based/external resources.

### 2.1.1 Machine Learning

Several studies have proposed approaches based on supervised and unsupervised learning. In (Bunschus et al. 2008) ontology generation approaches were discussed based on supervised learning whereas, Codo et al. (2007) worked on training a classifier for top 50 ambiguities from a mayo clinic, performed on clinical corpus. One of the drawbacks of using supervised learning methods is the requirement of huge training data with manually done labeling which ultimately increases time, cost and labor (Poesio et al., 2008). Unsupervised learning creates cluster to construct different hierarchies (Del et al., 2016). A study (Missisikoff et al., 2002) presented that unsupervised approach combined linguistic and statistics methods for performing ontology generation tasks for text but at the same time, it is followed by the drawback of dependency on statistical data without knowing the significance of the context.

#### 2.1.1.1 SQL Based Frameworks using Statistical Approach

The frameworks discussed here are SQL based using statistical approach. Different factors have been considered while comparing these frameworks, including testing and performance measures of the data. Wolfram Alpha, a famous search engine was developed by a team of researchers in Wolfram Research in 2009. It takes queries and requests submitted by users in the form of text fields and then performs computations and visualizations from structured data of knowledge base coming from different books and sites. At the end it displays results and interpretation of an input (Jonas, 2017).

An effort has been made by researchers to develop a framework which transforms English language input to SQL for the sake of relevant information retrieval from relational databases (Rao et al. 2010). The proposed framework provides the natural language to database queries translating infrastructure. However, the translation scope was limited to a user defined data dictionary containing most of the words to be used by the system. This framework allows users to extend by adding new translating grammar rules and data dictionary. For example, it has employed linguistic understanding with parse tree, and further maps the proximity of the patterns for the certain database concepts. The shortcoming of this system is that it does not support dialogue-style querying.

Ganti et al. presented a framework "Keyword++" to improve the existing tools to translate a keyword query to SQL statement (Ganti,2010). Proposed framework maps the query keyword to predicate and generate differential query pair (DQP) against the keyword, then measures the correlation between DQP and the predicates. Keywords to predicates mapping is further improved by aggregating the correlations which are measured on multiple query pairs extracted from a query log. A materialized mapping has been performed on the generated DQP from query log to translate query keywords to equivalent SQL statements. Proposed system has been tested on an entity table comprised of 8,000 laptops. Overall 0.1 million web search queries were extracted and trimmed to 500 queries as sample test set. Approximately, 2,000 keywords were extracted where each keyword had 41 DQP and took 1.61 seconds to compute the mapping. Experiments conducted on Keyword++ framework show that the effectiveness of the system is more the 80% compared to existing approaches.

Data sharing among various organizations could help to facilitate the evidence-based treatment by incorporating evidences from heterogeneous hospitals datasets Healthcare researchers and clinicians require tools to extract relevant information from clinical information system's data (Malik, 2018). These distributed databases contain different data design models e.g., Entity Relationship and Entity Attribute Value. Safari et al. proposed an algorithm to translate Restricted Natural Language Query (RNLQ) to SQL (Safari,2014). Generic algorithms have been used for mapping and translation. In the first step query terms are mapped to RNLQ via CliniDAL (Clinical Data Analytic Language) interface. Next temporal expression of the query is interpreted via a 2-layer rule based technique. Translation from RNLQ to SQL is performed via Top-K algorithm on the base of similarity that is further utilized by CliniDAL for the mapping process. The implemented prototype was tested on four categories of queries and it achieved 84% of accuracy.

Li et al. presented an approach which deals with complex input queries of multiple domains to translate them into SQL queries in a generic way (Li and Jagadish, 2014). The resulting SQL statements include query nesting, query joins, and query aggregation. A system has been developed based on the proposed approach named as NaLIR (Natural Language Interface for Relational databases), which incorporate these characteristics. The system reuses previous SQL statements from the query log to save query computation time.

TiQi, a natural language interface, allows to pose speech and text based queries in natural language (Lin, 2015). It is a web based tool and especially designed to access project's data. TiQi accepts user query and generates Traceability Information Model (TIM) which displays underlying object classes and attributes. TIM is stored in a centralized location to map unique nodes to access and specify data demanded by input query. In order to produce an up-to-date SQL output, H2, the JAVA SQL database, has been designed. This database engine provides support for data sources ranging from Jira to Excel Spreadsheet.

Palakurthi et al. presented a framework which classifies explicitly defined attributes present in a natural language query to convert them into various SQL clauses (Palakurthi, 2015). A statistical classifier CRF (Conditional Random Field) implemented to classify these attributes. The system has been tested on three domains (Academic, Restaurant, and Geo-query) and it has achieved accuracy of 70% and an F-measure of

85%.

Sujatha et al. has developed a system using the EFFCN algorithm, which used both semantic and syntactic knowledge to build an accurate match of input query to corresponding SQL query (Sujatha and Raju, 2016). It was tested on CPVbase with the precision and recall ranging 84%. Ontology merging and enhanced parsing process helps the system to prune the query for the sake of desired information. The authors suggested that future growth of NLIDB systems will be achieved via neural networks, machine learning and statistical parsing techniques, tackling abbreviated queries and dealing with temporal logic based complex natural language queries.

Mvumbi et al. proposed a system "NALI" to translate NL queries to SQL queries. It has been especially designed to address the portability issue of NLIDB (Natural Language Interface to Database) from one domain to another without customizing the tool manually and automatically generating the configuration model for the new domain (Mvumbi, 2016). The proposed approach reduced the manual workload to customized NLIDB. They introduced two authoring schemes (Top-down and Bottom-up) for customization in order to evaluate the best. Top-down approach, pre-harvests key lexical terms by using un-annotated sample NL queries. Furthermore, it includes semantics for negative form of verbs, comparative and superlative form of adjectives to reduce configuration workload. While the Bottom-up approach utilized database schema and data dictionary to generate configuration model automatically. The proposed system has been tested on Geoquery corpus, and revealed that top-down authoring approach results are much better compared to bottom-up for customizing a NLIDB system.

Sukthankar, et al. has presented a system to deal with simple as well as complex queries (Sukthankar, et al 2017). The proposed work focused on aggregate function, WHERE clause conditions and advanced clauses such as 'Having' and 'Order by'. The proposed system works well for single input query. The authors suggested to enhance the system by accepting multiple sentence queries and translate them into one resulting SQL query.

Seq2SQL framework inefficiencies (generalized to unseen schema and serializbility) and its Seq-to-Seq model has been improved with the help of new approach i.e. sequence-to-set based model proposed by Xiaojun (Xu,2017). The model has been implemented in "SQLnet" tool using Seq2SQL as baseline framework but eliminates reinforcement learning. Similar to SQLizer, a sketch based scheme has been implemented to parse the NL query, but each sketch has a dependency graph to predict the new sketch via using previous prediction of sketch. This new model improves the Seq2SQL results from 9% to 13% on various metrics.

Table 1 shows feature-based comparison of frameworks using statistical approach.

**Table 1** Features Comparison of SQL-based Frameworks Using Statistical Approach

| System Name | Language Support | Heuristic Rule Support | Interoperability | Usability Reported | Corrected Reported | Support Complex Querying | Performance Evaluation |
|---|---|---|---|---|---|---|---|
| **Wolfram Alpha 2009** | English to Wolfram Query Similar to SQL | Yes | No | Good | Good | Yes | Symbolic Computation, Knowledge Base, Ontology |
| **Keyword++, 2010** | English to SQL | Yes | No | Good | Good | Yes | Tested on 500 web queries and achieved >80% precision and recall |
| **RNLQ-SQL, 2014** | ClinDal Queries to SQL | Yes | Yes | Good | Good | Yes | Tested on RPAH-ICU Corpus, Accuracy 84% |
| **NaLIR, 2014** | English to SQL | Yes | Yes | Good | Good | Yes | Microsoft Academia Search(MAS) dataset, Good Recall and Precision |
| **TiQi, 2015** | NL Trace Query to SQL | Yes | Yes | Good | Good | Yes | Tested on Isolate and Easy Clinic datasets, Accuracy of supported and unsupported queries are respectively 92.6 %, 82.9 % |
| **Ashish, 2015** | English to SQL | No | Yes | Fair | Fair | No | Statistical classifier CRF trained on manually prepared data, and compare to Academic, Restaurant, Geo query, 70 % |
| **EFFCN, 2016** | English to SQL | No | No | Good | Good | Yes | Tested on CPVBase, Precision & Recall 84% |
| **NALI, 2016** | English to SQL | Yes | Yes | Good | Good | Yes | Tested on Geo query corpus, Customization, auto generation of configuration model |
| **nQuery, 2017** | English to SQL | Yes | No | Fair | Fair | Yes | Intelligent Table & attribute mapping, clause tagging aggregate function, group by and having clause |
| **SQLnet, 2017** | English to SQL | No | Yes | Good | Good | Yes | Improve Seq2SQL results from 9% to 13% on various metrics |
| **SQLizer, 2017** | English to SQL | Yes | No | Good | Good | Yes | Tested on MAS, IMDB and YELP databases, Accuracy 90% |

### 2.1.1.2 SQL Based Frameworks using Symbolic Approach

Alessendra and Alessendro (2012) proposed a framework and tested with 800 questing datasets about geo queries. They combined the through rules with weighing scheme that provides a ranking list of all selected candidates queries in SQL. Natural Language web Interface for Database (NLWIDB) is another commercial framework available to explore different databases (Rukshan et al., 2013).

A hybrid approach has been proposed to build a framework "NLKBIDB" using the methodology of NLIDB (Natural Language Interface to Databases) and KBIDB (Keyword Based Interfaces to Databases) by (Axita,2013). The author has explained various system agents that first accepts the NL query and passes it to various analyzers such as lexical, syntax, and semantic. If the query syntax is valid, the analyzed input further passed to the next agent is in the form of a tree structure. Tokens are mapped to generated knowledge base, if tokens are found in the knowledge base then a pointer is sent to the SQL generator, otherwise, the user is notified to reform query. This framework utilized the logical and conceptual schema of a database as the knowledge base. It is derived from the metadata of the database and knowledge experts help to update it. It has been tested on an agriculture survey database and reports 53 % accuracy against syntactically incorrect queries. Another commercial framework, developed in 2014, NQL which is widely used in organizations like university's databases. The purpose of NQL is querying database in natural language (Hessa and Emad, 2016). It parses English language queries and converts them into SQL. Furthermore, NQL algorithm was used in its implementation.

In 2015, another commercial framework for natural language to SQL was developed named as TR Discover (Dezhao et al., 2015). TR Discover is mainly use in domains like Life Sciences and law. It provides suggestions for construction of questions that belongs to natural language. TR Discover inherits SPAROL and SQL characteristics, that is the reason it uses feature based grammar and translation along with parsing. Table 2 displays the feature-based comparison of frameworks using symbolic approach.

**Table 2** Features Comparison of SQL-based Frameworks Using Symbolic Approach

| System Name | Language Support | Heuristic Rule Support | Interoperability | Usability Reported | Corrected Reported | Support Complex Querying | Performance Evaluation |
|---|---|---|---|---|---|---|---|
| **Alessandra, 2012** | English to SQL | Yes | No | Good | Good | Yes | Tested on GEOQuery Corpus 800 questions, Recall 88 % Accuracy 81% |
| **NLKBIDB, 2013** | English to SQL | Yes | No | Fair | Fair | No | Tested on Railway, college domain dataset, Solved 53% of syntactically incorrect queries % |
| **NLWIDB 2013** | Language Processing | Yes | No | Fair | Good | Yes | Language Processing |
| **NQL 2014** | NQL algorithm | Yes | No | Good | Fair | Yes | NQL algorithm |
| **TR Discover 2015** | Feature based Grammar. Parsing, FOL translation | Yes | No | Good | Fair | Yes | Feature based Grammar. Parsing, FOL translation |

## 2.1.1.3 SQL Based Frameworks using Connectionist Approach

Gulwani presented an auto synthesizing programming system "NLyze" to extract information from spreadsheet data without interacting with spreadsheet programming (Gulwani ,2014). The proposed system includes a Domain Specific Language (DSL) to deal with algebra of map, filter, and join etc. Compositional and Typed nature of DSL effectively translates the NL query and provides appropriate abstraction to the unskilled user. Translation of the NL query to spreadsheet programming is performed by a dynamic programming based algorithm, which converts the NL query into a ranked set of likely programs. The proposed algorithm combined two ideas- keyword programming and semantic parsing. Keyword programming approach has high recall but low precision, while semantic parsing has low recall and high precision. NLyze is specific to domain, purely based on typed synthesis, and targets only spreadsheets. In contrast, SQLizer, a similar tool to NLyze, provides domain independency by auto generating configuration model for new domain and target relational databases.

An intelligent agent based framework for databases to transform simple text to equivalent SQL query has been developed and presented in (Hessa et al., 2016). The authors have focused on parsing of extracted keywords via syntactic and semantic parser. Rules have also been designed for parser to learn the knowledge hidden in the natural language query. They used tools including Sphinx for speech recognition, MySQL as RDBMS, Stanford parts of speech (POS) tagger as syntax parser, Stanford named entity recognizer (NER) as semantic parser and for parsing complex queries ClearNLP. These tools are integrated in Intelligent Agent (IA) system which is implemented in Java. This system reported 80% accuracy in their test suite.

Inherent ambiguity of the NL query bound the underlying synthesizer of NLIDB system to automatically generate SQL representation. The scope of NLIDB is also limited by database agnostic (configuration required for every new database). A novel technique has been introduced by Yaghmazadeh et al. to automatically synthesize SQL queries and auto-generate a configuration model for new databases (Yaghmazadeh, 2017). Two ideas (typed-directed synthesis from NLP and repair technique from programming language) have been merged in this technique. At first stage semantic parser used to translate an NL query into a skeleton (a query sketch) which only represents the shape of query instead of full content. It does not generate SQL query by training the NL query on a specific database during semantic parsing. Initial sketch further needs to be refined because it does not capture the desired structure of input query, therefore it has been repaired via fault localization and database of repair tactics. Skeleton contained holes which are overcome by type-directed approach and converted to a complete SQL query and on each completion a confidence score assigned to query on the base of schema and contents of database.

An intelligent user interface minimizes the communication gap between user and the system. Adding an intelligent layer into the system eases the process of transforming the NL query to SQL statements. Singh et al. presented an intelligent NLIDB system named as "NLTSQLC" User, it is purely based on metadata and semantics sets for attributes and tables (Singh, 2016). This system takes input as the NL query, which is further processed for lower case conversion, tokenization, escape word removal, and part of speech tagging. System further classifies tagged tokens into relation, attributes and clauses. Finally, the system removes ambiguous attributes with the same name to generate the final SQL representation. Another framework, NLP Interchange Format (NIF) presented by Sebastian et al., (2012) works in the domain of semantic web. One of the major advantage of using NIF is, it provides interoperability that is actually global between different NLP tools. Kueri 2013, constructs query in faster manner. It facilitates user on a single click by getting their queries by simply typing search box (Yossi, 2016).

Zhong et al. proposed "Seq2SQL", a neural network-based framework for making an interpretation of natural language queries to relating SQL (Structured Query Language) representation (Zhong, Xiong, and Socher 2017). The proposed system reduces resulted query space and improves the execution accuracy of a system. This framework has utilized Reinforcement Learning (RL) rewards and cross entropy loss iteratively on query execution over the database to take in an approach to create unordered parts of the question, which are less reasonable for advancement by means of cross entropy misfortune. They released WikiSQL, a dataset of 87673 hand-explained cases of inquiries and SQL questions distributed over 26521 tables from Wikipedia. By applying strategy-based reinforcement learning (RL) with an inquiry execution condition to WikiSQL, Seq2SQL beats the best in class semantic parser by Dong and Lapata (2016). Utilizing the structure of SQL queries allows Seq2SQL to further reduce the output space of SQL queries, which leads to higher performance than Seq2Seq and the pointer model. Limiting the output space leads to more accurate conditions. Augmented pointer model generates higher quality WHERE clause conditions. Incorporating structure reduces invalid queries from 7.9% to 4.8%. Arimo was founded in 2012 with the purpose to support business intelligence and data science domain. Its user interface is inclined towards natural language processing that learns behavior patterns from big data. Similarly, Quepy 2012, allowed implementation of NLIDB systems using python language. It deals with complex queries and generate accuracy representation of data because it is based on NLTK framework (Jonas, 2017). In 2017, a natural language domain framework named as in2SQL was developed to deal with any natural language (Jeremy and Shashank, 2017). Similarly, Easy Query Building (EQB) 2017 is freely available that allows user queries in friendly way. All user requests can be described visually in natural language (EasyQuery, 2017).

To address the shortcomings of the NaLIR systems, such as tackling paraphrasing and various linguistic variations, another framework namely "DBPal" was presented in (Basik, 2018). To translate queries, it uses a novel translation model along with a feedback-based learning and auto-completion model to assist users in paraphrasing the partial query while formulating the database query. This system has shown significant accuracy improvement to build the complex queries. There are many commercial frameworks that uses connectionist approach to develop. Some of them are listed here. In 2011, The Needle framework was introduced. It was developed as website API for ecommerce websites and largely facilitates in shallow text analysis. It is also recognized as one of the fast search ecommerce API. Thoughtspot 2012, is available for data science and business domains. It has different significant features including ease of use, fast execution, minimal backlog maintenance and others (Ryan, 2018).

Table 3 has summarized all the features comparison of connectionist approach used in different frameworks.

**Table 3** Features Comparison of SQL-based Frameworks Using Connectionist Approach

| System Name | Language Support | Heuristic Rule Support | Interoperability | Usability Reported | Corrected Reported | Support Complex Querying | Performance Evaluation |
|---|---|---|---|---|---|---|---|
| **The NEEDLE 2011** | English to filter for SQL | Yes | Yes | Good | Good | Yes | API for shallow text analysis, Fast search |
| **Thoughtspot 2012** | English to SQL | Yes | Yes | Good | Good | Yes | Ecommerce API, Easy to Use, Faster execution, Reduce backlog, maintenance, Zero optimization to tune performance |
| **Arimo 2012** | English to SQL | Yes | Yes | Fair | Good | Yes | learn behavior patterns from Big Data |
| **Quepy 2012** | English to SQL, MQL, SPARQL | Yes | Yes | Good | Fair | Yes | deal with complex queries and generate accurate representation, open source, better for domain specific database |
| **NLP Interchange Format (NIF) 2012** | NLP tools output to RDF | Yes | Yes | Fair | Good | Yes | Global interoperability between different Natural language processing tools |
| **Kueri 2013** | English to SQL, and JSON for NoSQL | Yes | Yes | Good | Fair | Yes | Faster construction of query, efficient spelling checker, efficient data ambiguity handler, auto-completion of query |
| **NLyze, 2014** | English to SQL | Yes | Yes | Good | Good | Yes | Test on four different spreadsheet and achieved 98.2% accuracy and precision |
| **INLIDB, 2016** | English to SQL | Yes | No | Good | Good | Yes | Accuracy 80 % |
| **NLQTSQLC, 2016** | English to SQL | Yes | No | Good | Good | No | Tested on Manually prepared Dataset of questions, Accuracy 86%, Recall 80%, Precision 89% |
| **Seq2SQL, 2017** | English to SQL | No | No | Fair | Fair | Yes | Tested on WikiSQL Dataset, Execution Accuracy 59.4% and |

| | | | | | | | Logical form accuracy 48.3% % |
|---|---|---|---|---|---|---|---|
| **ln2sql 2017** | SQL | No | No | Fair | Fair | No | It can deal with any natural language |
| **Easy Query Builder (EQB) 2017** | English to SQL | No | No | Good | Fair | No | Visual representation of output |
| **DBpal, 2018** | English to SQL | No | No | Good | Fair | No | Tested on geographical data sets of United States |

### 2.1.2 Knowledge Based/ External Resources

In recent studies knowledge based approaches have been proposed to automate the ontology construction. A study presented by Harris et al. (2015) combines NLP and knowledge base for raw text ontologies. The approach was based on predefined dictionary of disorder type concepts that are expected to occur in the text. The drawback of this approach was increased labor cost for dictionary construction, the dependency of domain and limitations of patterns. Another work, presented by Cahyani et al. (2017) also focused on utilizing knowledge based on controlled vocabulary and data linked with corpus. Text2Onto tool was also used in this study for filtration on dictionary methods. For understanding of final concepts and candidate's relations, the work was also linked with pattern mapping of data. The limitation of this approach was it requires predefined relations to the domain and it semantic meaning is also not considered. Qawasmeh et al. (2018) proposed a work containing bootstrapping which was semi-automated involving preprocessing of manual text with extraction of concept. But the drawback involved in this approach was domain dependency on experts and involvement of labor in the process of development. Bhatia et al. (2018) another researcher focused on automating ontology generation of web pages that are retrieved. An et al. (2018) developed another approach that helped to transform database schema into automatically generated ontology. This was done with the help of crafted rules but like other approached, this approach has some serious problems. It required a predefined databased schema.

## 2.2 Translating Natural Language to Non Structured Query Language (NoSQL) Frameworks

In section 2.2 Natural Language to NoSQL frameworks have been presented based on two different matching approaches namely pattern matching and semantic matching, in context of rule based and syntax analysis, and syntactic pattern.

### 2.2.1    Rule Based & Syntax Analysis

Rule based and syntax analysis is a manual approach to set of rules that are formed for the representation of knowledge. This representation involves the decision to conclude various scenarios. A study presented by Abacha et al. (2011) shows that medical entities and the relationship of medical text and rule based syntactic pattern are basically semi automatically built according to the criteria of semantics from a corpus. Similarly, Ono et al. (2001) defined an approach for extraction of protein interaction information from the literature presented in dictionary named rule-based and syntax based analysis. In the said approach protein to protein interaction was presented.

#### 2.2.1.1  NoSQL frameworks Using Semantic Matching

Semantic web contained a large number of linked open data repositories. Due to the complex nature of SPARQL queries, it is difficult to formulate these by a naive user or even by an expert. Bretonnel et al. developed a prototype system Linked Open Data Question Answering (LODQA) to transform plain text queries into corresponding SPARQL statements (Kim and Cohen 2013). Techniques used to implement the prototype version included parsed via Enju, pattern based matching using chunks of base noun, targeting performed via pattern matching, shortest path find via Dijkstras algorithm, ontology searching performed via OntoFinder, and to determine and selection of a predicate as default. Some modules of the proposed system do not perform as desired, so these are configured, and an integration testing is performed for all modules in future.

To explore data from these domains, Karim has developed an efficient tool "Sem-QAS" (Karim et al. 2013). The main function of this tool is to convert natural language questioning into corresponding SPARQL query through identification of unique atomic constraint and their relation present in the input question. This tool generates and combines triple patterns to output complex SPARQL queries for atomic constraint. Recall and precision of the system mainly measured for association operators and scope modifiers processing. It is tested on Mooney Job corpus for correctness and efficiency (Karim et al. 2013).

To convert natural language queries to non-SQL database, there is a need to develop systems that can interpret non-English languages. In this regard, a system for Arabic language named as "AR2SPARQL" has been developed by (Al Agha and Abu-Taha, 2015) to enable non-technical users to query RDF graphs. Query ambiguity is resolved via linguistic as well as semantic approach. System has been tested on two corpora and showed good performance statistics for precision and recall. Another article which uses Arabic language as a case study for Natural Language Interface for Relational Database is presented in (Hammo, Abu-Salem and Lytinen, 2002).

Exploring graph databases via natural language query also has strong potential. There has been a little amount of inter pattern technique and translating them into corresponding SPARQL query. A framework for real life application related to organic farming will be the target in the upcoming work. A large amount of work has been carried on querying ontologies and RDF data via English language queries. A natural language to SPARQL querying framework has been introduced by (Sæbu, 2015). The proposed system does not demand background knowledge to build a query. The C-system analyzes an input information request and generates a SPARQL query against it to explore required information from databases. Job searching is a common task for unemployed persons. Data available on job search domains is annotated semantically.

Information requested by user in his own words can be expressed without confining to point, click, scroll or search to choose correct class and features. To ease the task of GraphAware NLP is one of those available tools, which have been presented by (albertodelazzari et al. 2014). Developed as a plugin for Neo4j graph database, GraphAware NLP provides a group of tools in form of procedures, APIs and background process. Accurately converting a plain text question into corresponding database statement is the main goal of NLIDB domain. There has been an effort to develop a system named as MANTRA QA by (Oro and Ruffolo, 2015). It transforms a plain text query into SPARQL and Cypher statement. It is a mixture of grammar and logic-based concepts to accurately find out the concepts and relations in specific knowledge domains. Primarily, it has been tested on tourism and finance domains benchmarks.

Table 4 shows NoSQL and graph databases frameworks using semantic matching.

**Table 4** Features Comparison of NoSQL-based Frameworks Using Semantic Matching

| System Name | Language Support | Datasets Supported | Domain Supported | Performance Evaluation |
|---|---|---|---|---|
| **LODQA, 2013** | English to SPARQL | SNOMED CT | Life Sciences | Not Available |
| **SemQAS, 2013** | English to SPARQL | Mooney Job | Job Search | Precision 100% and recall 99 % |
| **AR2SPARQL, 2015** | Arabic to SPARQL | OWL ontology based on United State Geography data | Question answering system for Quran | Test on geography data set and achieved precision 88%, Recall 61%, F-Measure 0.72 while for disease data set 82%, 62% and 0.71 |
| **OptiqueNLQF, 2015** | English to SPARQL | NPD Ontology | Petroleum companies | Not Available |
| **MANTRA QA, 2015** | English to SPARQL & Cypher | Manually Prepared questions | Tourism | Not Available |

#### 2.2.1.2 NoSQL frameworks Using Pattern Matching

Semantic Web Interface Using Pattern (SWIP) has been presented by Camille et al. (2013). In this framework French ecology and agriculture is mainly focused. Furthermore, SWIP uses English language to SPAROL and dataset named QALD-3 supports this framework.

Table 5 represents NoSQL and graph databases frameworks using pattern matching.

**Table 5** Features Comparison of NoSQL-based Frameworks Using Pattern Matching

| System Name | Language Support | Datasets Supported | Domain Supported | Performance Evaluation |
|---|---|---|---|---|
| **SWIP, 2013** | English to SPARQL | QALD-3 | French ecology and agriculture | Tested on Music brains dataset achieve 51% precision, recall, and F-measure |

### 2.2.2 Syntactic Pattern

Syntactic pattern is a well-known approach in the area of natural language processing especially ontology engineering and extraction from data (Maynerd et al., 2009). Unlike other approaches, this approach consists of huge amount of crafted syntactic patterns therefore, it has high recall and low precision. And this makes it domain dependent also (Reiss et al., 2008).

Hearst in 1992 extracted hyponymy lexical relations from text based hand written patterns and parts-of-speech were used as tags in it. In another study by Downey et al. (2004), the approach was used to learn the patterns from text and extract information from them. Text2Onto as previously discusses, combines other approaches like machine learning with basic linguistic approaches and used in POS tagging. The drawback of syntactic pattern is the limit number of available patterns and dependency of domain. There are also some major issues of scalability, domain knowledge and labor.

Tiddi et al. (2012) presented an approach. The purpose of the approach was to generate web content via syntactic patterns for the relations that exist in linked open data. Furthermore, it was used to extract new entities from web and ontology construction. It has also some limitations like its limit to RDF scheme only and need of domain of interest for input.

### 3. Discussion

The aim of this survey is to explore the state of the art natural language querying frameworks for databases. These natural languages to database querying frameworks have a four-decade long history and several efforts have been made to facilitate end users. Initial tools were designed to deal with databases on the small scale and those systems were not scalable for commercial use. With advancement in technology, new commercial natural language to database querying frameworks were developed supporting multiple databases. In the literature, related studies and surveys have been found which focus particularly on the natural language to database querying frameworks and most of the available surveys are outdated.

We have included frameworks developed around 2008 to 2018 and found 47 frameworks related to our topic. Out of 47, we have selected 35 frameworks closely relevant to our targeted topic for survey. A feature comparison table has been compiled for those tools which are available for commercial purposes. Initial frameworks which were developed for translation do not fulfill the definition of current databases due to the heterogeneous nature of data, and these tools were only designed to deal with small scale databases. These frameworks have been categorized according to structured and unstructured databases. Multiple query languages are used to retrieve information from databases, but we focused only on the popular query languages of today. Furthermore, we have presented reviews of frameworks which translate natural language query to SQL, MQL, SPARQL, RDF, CYPHER, and GREMLIN for both structured and unstructured databases.

Lastly, we have found that 70% of the work in the natural language to database querying is carried out for SQL. The share for NoSQL languages such as for SPARQL, CYPHER, and GREMLIN are 15%, 10% and 5% respectively. With the increasing popularity of NoSQL especially graph databases, we urge researchers to focus on developing new natural language to database frameworks for CYPHER and other graph languages. It has also been observed that most of the available natural language to database querying frameworks support English language only. Few efforts have been reported where researchers have worked on Portuguese and French. Multi-language support or dedicated systems in international languages are desirable to make the overall data-driven process easy.